\documentclass[usenatbib]{mn2e}
\usepackage{graphicx}
\newcommand{\text}[1]{\quad\mbox{#1}\quad}

\newcommand{\sub}[1]{_{\mbox{\tiny #1}}}
\newcommand{\bfrac}[2]{\left(\!\frac{#1}{#2}\!\right)}

\newcommand{\apj}{ApJ}
\newcommand{\apjs}{ApJS}

\newcommand{\aap}{A\&A}

\title{Supercollapsars and their X-ray Bursts} 
\author[S.S. Komissarov and M.V.Barkov]
{Serguei S.~Komissarov$^{1,3}$ 
\thanks{ serguei@maths.leeds.ac.uk (SSK)} and  
Maxim V.~Barkov$^{1,2,3}$
\thanks{E-Mail: bmv@maths.leeds.ac.uk (MVB)}\\
$^{1}$Department of Applied Mathematics, The University of Leeds,
Leeds, LS2 9GT, UK\\
$^{2}$Space Research Institute, 84/32 Profsoyuznaya Street, Moscow
117997, Russia\\
$^{3}$NORDITA, AlbaNova University Centre, Roslagstullsbacken 23, SE-10691 Stockholm, 
Sweden}
\begin{document}
\date{Received/Accepted}
\maketitle
                                                                              
\begin{abstract}

The very first stars in the Universe can be very massive, up to $10^3M_\odot$. 
If born in large numbers, such massive stars can have strong impact on the subsequent  
star formation, producing strong ionising radiation and contaminating the primordial gas      
with heavy elements. They would leave behind massive black holes that could act as seeds 
for growing supermassive black holes of active galactic nuclei. Given the anticipated fast
rotation, such stars would end their live as supermassive collapsars and drive powerful 
magnetically-dominated jets. In this letter we investigate the possibility of observing 
the bursts of high-energy emission similar to the Long Gamma Ray Bursts associated with 
normal collapsars.  We show that during the collapse of supercollapsars, the 
Blandford-Znajek mechanism can produce jets as powerful as 
few$\times10^{52}$erg/s and release up to $10^{54}$erg of the black hole rotational 
energy. Due to the higher intrinsic time scale and higher redshift the initial bright 
phase of the burst can last for about $10^4$ seconds, whereas the central engine would 
remain active for about one day.  
Due to the high redshift the burst spectrum is expected to be soft, with the    
spectral energy distribution peaking at around 20-30keV. The peak total flux density is 
relatively low, $10^{-7}\mbox{erg}\, \mbox{cm}^{-2} \mbox{s}^{-1}$, but not prohibitive.   
If one supercollapsar is produced per every mini-halo of dark matter arising from 
the 3-$\sigma$ cosmological fluctuations then the  whole sky frequency of such bursts could 
reach several tens per year.      
 
\end{abstract}

\begin{keywords}{
cosmology: early Universe -- black hole physics -- gamma-rays: bursts -- X-rays: bursts --
radiation mechanisms: non-thermal -- relativity -- magnetic fields}
\end{keywords}

\section{Introduction}
\label{intro}

According to the modern hierarchical clustering theories of galaxy formation 
the first stars are born within collapsed haloes of dark matter of $\simeq 10^6M_\odot$ 
at $z\simeq 20$. The primordial gas falls into the potential well of these 
haloes and fragments into clamps of $\simeq 10^3M_\odot$ via gravitational instability 
\citep{BCL02}.  Because this gas is metal-free its cooling is rather slow and further 
fragmentation into smaller clamps seems to be avoided \citep[cf.][]{TAS09,SGB09}. 
Instead, the clumps contract in a quasi-static fashion as a whole, suggesting that 
the first stars can be very massive indeed, $M>100M_\odot$. 
The actual initial mass function (IMF) of first metal-free stars (Population III stars), 
however, is not known yet as too many factors come into play, making the problem 
intractable analytically, and rather challenging numerically.
In particularly, the initial mass of protostars can be very small, down to 
$10^{-3}M_\odot$, and the eventual accumulation of mass proceeds via accretion 
of the surrounding gas. Very high accretion rate, 
$\dot{M}>\dot{M}_c \simeq 4\times10^{-3} M_\odot\mbox{yr}^{-1}$, may limit the final 
mass to few hundreds solar masses as the protostellar luminosity reaches the Eddington limit 
\citep{OK03}. For lower accretion rate, the accretion may proceed even after the onset of 
nuclear burning in the stellar core and result in the final mass $M\simeq 10^3M_\odot$.   
Numerical studies of cosmological gravitational instability suggest that, although in principle 
the accretion rate can be as high as few$\times 10^{-2}M_\odot\mbox{yr}^{-1}$, 
in reality the rotational support 
against gravity often become important and reduces the rate below $\dot{M}_c$ \citep{G07}.       
\citet{Ohk09} studied the evolution of accreting Population III stars from the pre-main 
sequence evolution to the core-collapse and confirmed that the final mass can be as 
large as $10^3 M_\odot$. Very massive first stars are also predicted in theories involving
dark matter annihilation \citep[e.g.][]{Nat09}. 

Population III stars with masses $140M_\odot \le M \le 260M_\odot$ most likely end their 
life as pair-instability supernovae which leave no compact remnant behind \citep{F01}. 
If such stars were the main outcome of initial star formation they would overproduce 
heavy elements in the early Universe, in conflict with the observations of extremely 
metal pure stars in galactic bulges and the observed abundances of intergalactic and 
intercluster medium \citep{UN02,HW02,CL02}.      

More massive 
stars, which will be referred to as Very Massive Stars (VMSs) are expected to collapse 
into black holes with very little mass loss \citep{F01}.   
They would leave behind massive black holes (MBHs), which could play the role of seeds 
for the supermassive black holes (SMBHs) of Active Galactic Nuclei (AGNs). 
Assuming that MBHs are formed at the rate of
one per mini-halo developed from a 3$\sigma$-fluctuation, \citet{MR01} estimated their 
density to be around 5000 per galaxy like Milky Way and their total mass comparable to 
the total mass of SMBHs. This suggests that SMBHs could form 
via mergers of MBHs, the idea that has being actively developed in recent years. 
Even more massive VMSs, with $M\simeq 10^6M_\odot$, could be formed in more massive dark 
matter haloes, with total mass $M\simeq 10^8M_\odot$, collapsed at $z\simeq10$ 
\citep{BL03,BVR06}. Although much more rare, such events can provide an alternative 
way of producing SMBHs.

From the observational perspective it is difficult to distinguish between 
a VMS and a cluster of less massive Population III stars.  This suggests to 
investigate the potential observational consequences of VMS collapse, which could
be quite spectacular because of the very high mass involved. Given their 
expected fast rotation, it seems likely that supercollapsars (classified as type-III 
collapsars in \citet{H03}) develop accretion disks, drive relativistic jets,  
and produce bursts of high energy emission in the fashion similar to their less 
massive relatives \citep{MW99,BK08}. If detected, they would become the most distant 
strong sources of light and provide us with a new way of probing the physical 
conditions in the epoch close to the ``Dark ages''.  Even non-detection could be
useful, allowing to set constraints on models of star formation 
in the early Universe and the origin of SMBHs. 
There has been a number of 
papers looking into cosmological evolution of GRBs, including those from the Population III
stars \citep[e.g.][]{BL02,LD07,NB07}. They assumed that the Population III GRBs  
are similar to those from the lower redshift Population II stars whose mass is 
significantly below $100M_\odot$. However, because of their very high mass and 
redshift, the GRB-like bursts of supercollapsars can be rather special. 

There are two crucial differences between a normal collapsar and a supercollapsar. 
One is that instead of a proto-neutron star of solar mass the supercollapsars
develop proto-black holes of tens of solar masses,  
within which the neutrinos from electron capture are trapped \citep{F01,Su07}. 
The other is that the accretion disks of supercollapsars are far too large and cool 
for the neutrino annihilation mechanism. This has already been seen in the numerical 
simulations of supercollapsar with mass $M=300M_\odot$ \citep{F01}. Utilising  
the study of hyper-accreting disks by \citet{Bel08} we find that at best the  
rate of heating due to this mechanism is 
\begin{equation}
\dot{E} \simeq 2\times10^{48} \dot{M}_0^{9/4}  M_{h,3}^{-3/2}  
 \mbox{erg s}^{-1},
\end{equation}    
where $\dot{M}$ is the accretion rate and $M_h$ is the black hole mass. 
(Here and in other numerical estimates below we use the following 
notation:  $\dot{M}_k$ is the mass accretion rate measured in the units of 
$10^k M_{\odot} \mbox{s}^{-1}$ and $M_k$ is the mass measured in the units of 
$10^kM_\odot$.) Such low values  
have lead \citet{F01} to conclude 
that the magnetic mechanism is the only candidate for producing GRB jets 
from supercollapsars. In the following we analyse one particular version 
of the mechanism, namely the one where the jets are powered by the rotational 
energy of the black hole via the Blandford-Znajek process \citep{BZ77,BK08}.    

\section{Blandford-Znajek jets from Supercollapsars}  
\label{BZ-jets}

VMSs are expected to rotate rapidly, close to the break-up 
speed and hence produce rapidly rotating MBHs. Moreover, in the  
absence of strong magnetic field in the prestine primordial gas, VMSs 
will be weakly magnetized and as the result could develop rapidly rotating 
cores \citep{WH06}. This suggests that the spin parameter of MBHs  
can be very high, $a \simeq 1$, yeilding the enormous rotational 
energy $E\sub{rot} \simeq 5\times10^{56} M_{h,3} \mbox{erg}$. 
In order to estimate the Blandford-Znajek luminosity we need to know the 
strength of magnetic field accumulated by the hole. The usual approach 
is to relate it to the gas pressure in the disk and for this we need to 
know the parameters of the accretion disk itself. Accurate determination 
of these parameters and their time evolution requires to know the structure 
of VMS prior to the collapse and the physics of accretion disk. Unfortunately, 
this information is lacking at the moment. In particular, although the structure 
of  supermassive stars, $M_s \gg 10^3M_\odot$ is very well described by a polytropic
model with $n=3$ \citep{ZN71}, we are more interested in stars with $M_s \leq 10^3M_\odot$. 
\citet{F01} studied the structure and evolution of a $300M_\odot$ star. Prior to the 
collapse the star entered the red giant phase and expanded from $R_s=4\times10^{12}$cm 
to $1.5\times10^{14}$cm. However, the initial rotation rate of this star was  
slow compared to about $\simeq50\%$ of the brake-up speed invoked 
the current single-star model of GRB progenitors. In this model, the progenitors 
remain chemically homogeneous and compact all the way up to the collapse 
\citep{YL05,WH06}.  Given this lack of information 
about the progenitor structure we will follow \citet{b90} and assume 
$\rho \propto R^{-3}$ ($R_c\leq\! R \leq\! R_s$) 
distribution of mass density in the star prior to collapse.    
In fact, this distribution agrees reasonably well with 
the numerical models of rapidly rotating low 
metallicity stars considered as likely progenitors of normal GRB
\citep[see Figure 2 in ][]{KNJ08}. As to the stellar rotation, we will assume 
that it is uniform ($\Omega$=const) in the stellar envelope,
with $50\%$ of break-up speed at the stellar surface.   

Due to the slow neutrino cooling, the accretion disks of supercollapsars are 
expected to be radiatively inefficient, with possible exception only for the very 
inner region. This suggests to use the ADAF (Advection Dominated Accretion Flow) 
model \citep{NY94} to describe these disks. Since the radiation pressure dominates,  
we can use the ratio of specific heats $\gamma=4/3$ which gives us 

\begin{equation} 
v\sub{in} \simeq \frac{3\alpha}{7}v_k, 
\qquad c_s^2 \simeq \frac{2}{7}v^2_k,
\qquad H \simeq R c_s/v_k,
\label{ADAF}
\end{equation} 
where $v\sub{in}$ is the accretion speed, $c_s$ is the sound speed,
$v_k=\sqrt{GM/R}$ is the Keplerian speed, $H$ is the vertical disk scale, 
and $\alpha$ is the effective viscous stress parameter of the $\alpha$-disk 
model \citep{SS73}.  
The disk density and pressure can be estimated combining the above equations with the 
expression for the mass accretion rate, $ \dot{M}\simeq2\pi R H\rho v\sub{in}$. 
Straightforward calculations yield 
\begin{equation} 
P \simeq \frac{\sqrt{14}}{12\pi\alpha} \frac{\dot{M}(GM)^{1/2}}{R^{5/2}}.
\label{p}
\end{equation} 
The poloidal magnetic field should scale with the thermodynamic pressure, so we 
write $B^2 = 8\pi P/ \beta$, where $\beta$ is the magnetization parameter. 
Applying this equation at the radius of the marginally bound orbit $R\sub{mb} = f_1(a)R_g$, 
where $f_1(a)=2-a+2(1-a)^{1/2}$ and $R_g=GM_h/c^2$ is the BH's gravitational radius,  
we find   

\begin{equation} 
B\sub{mb} \simeq 3\times10^{13}  f_1^{-5/4} \beta_{1}^{1/2} 
   \dot{M}_{0}^{1/2}  \mbox{Gauss},
\label{b}
\end{equation}
where $\beta_1=\beta/10$.\footnote{The inner edge of the disk varies between 
the marginally bound and the marginally stable orbits, depending of its thickness. 
In our calculations this does not make much difference.} 
If the magnetic field is generated in the disk then it 
is likely to change polarity on the scale $\simeq H$. This may lead to 
significant variation in the strength and polarity of the magnetic field 
accumulated by the black hole and reduce the Blandford-Znajek luminosity 
\citep[e.g.][]{BB09}. We will assume that this effect is accounted for in 
the value of $\beta$.       

The power of jet energized via the Blandford-Znajek mechanism can be estimated 
using the monopole solution for magnetospheres of rotating black holes 
\citep{BZ77}, which gives   
\begin{equation} 
L\sub{BZ} =
\frac{1}{3c}\left(\frac{\Psi_h\Omega_h}{4\pi}\right)^2
\label{bz1}
\end{equation} 
where $\Omega_h= f_2(a)c^3/GM_h$ is the angular velocity of BH, 
$f_2(a)= a/2(1+\sqrt{1-a^2})$, and $\Psi_h$ is the magnetic flux threading 
one of the BH's hemispheres. 
Inside the marginally bound orbit the disk plasma quickly dives 
into the BH and the magnetic flux can be roughly estimate as 
$\Psi=2\pi R\sub{mb}^2 B\sub{mb}$ \citep{RGB06}. Combining this result with 
Eqs.(\ref{b},\ref{bz1}), we find  
\begin{equation}
L\sub{BZ} \simeq
\frac{\sqrt{14}}{9}\frac{f_1^{3/2}f_2^2}{\alpha\beta} \dot{M}c^2 \simeq 
\frac{0.05}{\alpha_{-1}\beta_1}  \dot{M}c^2, 
\label{bz2}
\end{equation} 
where $\alpha_{-1}=\alpha/0.1$ ( for $0.5\!<a\!<1$ the combination $f_1^{3/2} f_2^2$ 
depends weakly on $a$ and is approximately $1/4$). 

The mass accretion rate can be estimated following the procedure described in 
\citet{BK09}. The total accretion time includes the travel time of the rarefaction 
wave send into the stellar envelope by the core collapse, the time of the 
envelope collapse, and the disk accretion time, which gives the largest 
contribution. Accounting only for the disk contribution, the 
accretion time scale for the stellar matter located in the progenitor at radius $R$ 
is $t \propto l^3/M^2$, where $l=\Omega R^2$ and $M(R)$ is the stellar 
mass enclosed within radius $R$. Then $\dot{M} = dM/dt \simeq (dM/dR)/(dt/dR)$, where  
for the Bethe's model we have $dM/dR \simeq M_s/(R\ln(R_s/R_c))$ 
and $dt/dR \simeq 6t/R$, where $R_c$ is the stellar core radius. 
Collecting the results we obtain 
\begin{equation}
\dot{M} \simeq \frac{1}{6}\frac{M_s}{\ln(R_s/R_c)} \frac{1}{t} 
\simeq 36 \frac{M_{s,3}}{t} M_\odot \mbox{s}^{-1},
\label{m-dot}
\end{equation}
where $t$ is measured in seconds and we used $R_s/R_c=100$.\footnote{ 
For a $300M_\odot$ star at $t\simeq 8$s this gives 
$\dot{M}\simeq 1.3M_\odot \mbox{s}^{-1} $, which agrees reasonably well 
with numbers given in \citet{F01}.} 
Here, we assumed that the whole of the disk is accreted by BH, following the 
original ADAF model. However, it has been argued that this model has to 
be modified via including disk wind  \citep[Advection Dominated Inflow Outflow 
Solution (ADIOS),][]{BB99}, which 
implies a mass loss from the disk and a smaller accretion rate compared to 
Eq.(\ref{m-dot}). While the arguments for disk wind are very convincing, 
the actual value of mass loss is not well constrained and can be rather 
low. Given Eq.(\ref{m-dot}) the power of BZ-jet is 
\begin{equation}
L\sub{BZ} \simeq
\frac{3.2\times 10^{52}\epsilon_m} {\alpha_{-1}\beta_1} \frac{M_{s,3}}{t_2} 
\mbox{erg}\,\mbox{s}^{-1},
\label{bz3}
\end{equation}
where $t_2=t/100$ and $\epsilon_m<1$ is the fraction of the disk mass reaching the BH. 
Given the jet propagation speed inside the star, $v_j\simeq 0.2c$, deduced from 
axisymmetric numerical simulations 
\citep{BK08}, the jet breakout time is expected to be around of few hundred seconds 
and, thus, the numerical factor in Eq.(\ref{bz3}) gives us the optimistic jet 
power at the time when it becomes observable. In fact, the initial 
influx of mass through the polar column is very large and activation of the 
Blandford-Znajek mechanism can be delayed \citep{KB09}. The very latest time for 
the activation is given by the free-fall time of the whole star,  
\begin{equation}
  t\sub{ff} \simeq 1000 R_{s,12}^{3/2} M_{s,3}^{-1/2} \mbox{s},
\label{free-fall}
\end{equation}
as by this time the polar column becomes completely empty. 

The total duration of the jet production phase has to be similar to the 
disk lifetime. If VMS is rotating at half of the break-up speed then the initial 
outer edge of the disk is at $R_d \simeq R_s/4$. Ignoring the edge expansion 
due to accumulation of angular momentum, the disk life time is given 
by its "viscous" time scale      
 
\begin{equation} 
t\sub{ce} \simeq \frac{2R_s}{3v\sub{in}(R_s)} 
          \simeq 5000 \alpha_{-1}^{-1}
     R_{s,12}^{3/2} 
     M_{s,3}^{-1/2} \mbox{s}, 
\label{cet}
\end{equation} 
where $R_{s,12}$ is the stellar radius measured in 
$10^{12}$cm. By this time the BZ power will be significantly reduced 
but could still play a role in shaping the light curve of afterglow emission
\citep{BK09}. 

Using the mass accretion rate given by Eq.(\ref{m-dot}) we can check if the neutrino cooling 
needs to be included in the model. Under the conditions of the supercollapsar's 
disk its cooling is dominated by pairs.  
Using the well known equation for this cooling rate \citep[e.g.][]{YKGH01} we can 
compare the cooling time with the accretion time at a given disk radius. The result is 

\begin{equation}
\frac{t_d}{t\sub{cool}} \simeq 0.3 
  \alpha_{-1}^{-9/4} (R/R_g)^{-13/8}
   \dot{M}_{-1}^{5/4}  M_{s,3}^{\!-3/2}.
\label{cooling}
\end{equation}  
Thus, except for the very 
inner part of the disk, the neutrino cooling is indeed inefficient. 

The high BZ power given by Eq.(\ref{bz3}) suggests that the GRB-like burst emission from 
such jets could be seen even from $z\simeq 20$ and in the next section we 
discuss the properties of such bursts in more details.     

\section{Observational signatures}
\label{signatures}

Assuming that the radiation mechanism of the supercollapsar jets is similar to that 
of normal GRB jets, we expect the peak in the spectral energy distribution of the prompt 
emission  in the source frame to be around  $0.5$MeV. However, the cosmological 
redshift effect reduces the peak down to   
\begin{equation}
E\sub{max} \simeq 25\mbox{keV} \bfrac{1+z}{20}^{-1}, 
\label{peak freq}
\end{equation}
which is still inside the energy window of {\it Swift}'s BAT. 
For the same reason the observed total duration of the burst increases up to  
\begin{equation}
t_b \simeq 1 \bfrac{1+z}{20} \alpha_{-1}^{-1}
     R_{s,12}^{3/2} M_{s,3}^{-1/2} \mbox{day}.
\label{duration}
\end{equation} 
The characteristic source frame time scale for the decay of BZ luminosity 
in the model presented above is given by the time since the onset of the collapse. 
Thus, the initial time scale for the burst decay will be of order of the jet break out 
time, few$\times 10^2$ seconds, or a bit longer if the activation of the BZ-mechanism 
is significantly delayed. In the observers frame this translates into 
few$\times 10^3 - 10^4$ seconds. Thus, these bursts would appear not only unusually 
soft but also unusually long-lasting.        

The total flux density of the burst emission received on Earth and the 
isotropic luminosity are related via $F = L/(4\pi r\sub{L}^2)$,
where $r\sub{L}$ is the luminosity distance to the source \citep[e.g.][]{P93}. However, 
the emission from GRB jets is highly anisotropic due to the relativistic beaming.  
Moreover, not all of the Blandford-Znajek power is converted into the radiation 
within the energy window of the receiver. This leads to 

\begin{equation}
   F = \epsilon_c\frac{L\sub{BZ}}{4\pi r\sub{L}^2 {\cal A}}, 
\label{flux1}
\end{equation} 
where ${\cal A}\ll 1$ is the solid angle of the radiation beam and 
and $\epsilon_c<1$ is the conversion efficiency.  In flat Universe 

\begin{equation}
   r\sub{L} = \frac{c}{H_0}(1+z) \int\limits_{0}^{z} 
   (\Omega_m(1+z)^3+\Omega_{\Lambda})^{-1/2} dz.
\label{lum dist}
\end{equation}
For $z=20$ and the density parameters $\Omega_{\Lambda}=0.72$,
$\Omega_{m}=0.28$ \citep{k09} this gives us

\begin{equation}
F \simeq  2 \times 10^{-7} 
         \frac{\epsilon_{c,-1} \epsilon_m}{\alpha_{-1}\beta_1 {\cal A}_{-3}} 
         M_{s,3} t_3^{-1}  
    \mbox{erg}\, \mbox{cm}^{-2} \mbox{s}^{-1},
\label{flux}
\end{equation}
where $\epsilon_{c,-1}=\epsilon_c/0.1$, ${\cal A}_{-3}={\cal A}/10^{-3}$, 
and $t_3=t/10^3$. One can see that for the first $10^4$s this is above the 
sensitivity of BAT, $10^{-8} \mbox{erg}\, \mbox{cm}^{-2} \mbox{s}^{-1}$, and thus 
such a burst could trigger BAT. Having said this, we keep in mind 
that there is a great deal of uncertainty with respect to the values of various 
parameters appearing in Eq.(\ref{flux}). 

The time dependence in Eq.(\ref{flux}) gives the evolution of mean bolometric 
flux. It is not clear if the supercollapsar bursts will also exhibit the fine 
substructure characteristic of normal GRBs. If the variability of normal GRBs 
is due to internal shocks in baryon dominated flow, as this is proposed 
in the currently most popular model of prompt gamma-ray emission \citep{MR94}, 
then the supercollapsar burst produced by magnetically-dominated BZ jet may well 
be smooth and featureless. However, there are models of normal GRBs that 
attribute the observed variability to unsteady magnetic dissipation 
\citep{LB03,GUB09,KN09}. If they are correct then the supercollapsar 
bursts will also show fine substructure. 

In order to estimate the observed rate of such burst we assume, following 
\citet{MR01}, that the dark matter mini-haloes that host 
supercollapsars arise from 3-$\sigma$ fluctuations that constitute only $\simeq 0.3\%$ 
of the dark matter matter of the Universe and that only one supercollapsar per mini-halo 
is produced. The total mass per Mpc$^3$ at $z=20$ is 
$M\sub{Mpc} \simeq 1.5\times 10^{15} M_\odot$. The number density of 3-$\sigma$ 
mini-haloes is then 
$$
   n\sub{mh} \simeq 0.003 \frac{\Omega\sub{dm}M\sub{Mpc}}{10^6 M_\odot} 
       \simeq 10^6 \mbox{Mpc}^{-3}. 
$$ 
Let us assume, for the sake of simplicity, that all supercollapsars go off 
simultaneously at cosmological time $t_e$ corresponding to $z=20$ (a moderate 
spread around this redshift will not significantly change the result). 
In flat Universe the observed time separation between events occurring 
simultaneously at $r_0$ and $r_0+dr_0$, where $r_0$ is the comoving radial 
coordinate, is $dt_o=cdr_0$. The corresponding physical volume within one steradian of 
the BAT's field of view, is 
$$
dV=a^3(t_e) r_0^2 dr_0,
$$ 
where $a(t_e)=(1+z)^{-1}$ is the scaling factor of the Universe at $t=t_e$  
(in the calculations we fix the scaling factor via the condition $a(t_o)=1$). 
$r_0$ and $t_e$ are related via $r_0=r\sub{L}(1+z)^{-1}$. Putting all this together 
we find the rate to be 
$$
   f_c = {\cal A}\frac{ c n\sub{mh} r\sub{L}^2}{(1+z)^5} 
   \simeq 4\, {\cal A}_{-3} \bfrac{n\sub{mh}}{10^6} \mbox{yr}^{-1} \mbox{sr}^{-1}.  
$$     
Recent high-resolution simulations of cosmological star formation indicate the 
possibility of further fragmentation of gas clumps in minihaloes, resulting in formation 
of binary or even multiple protostars in some realizations \citep{TAS09,SGB09}. 
Thus, the theoretical rate of VMS formation can be significantly smaller compared 
to the one used in our calculations, making the supercollapsar bursts rare events.       
This may explain why such bursts have not been seen so far. 

\section{Conclusions}

In spite of the significant progress in the astrophysics of Gamma Ray Bursts, both 
observational and theoretical, it may still take quite a while before we fully understand 
both the physics of the bursts and the nature of their progenitors. At the moment there 
are several competing theories and too many unknowns. Similarly, we know very little about 
the star formation in the early Universe.  For this reason, the analysis presented above 
is rather speculative and the numbers it yields are not very reliable. 
Further efforts are required to develop a proper theory of supercollapsars and 
to make firm conclusions on their observational impact.
On the other hand, our estimates suggest that if we are on the right track then 
the X-ray bursts of supercollapsars may be detectable already with {\it Swift}. 
The expected very long duration of bursts and their relatively low brightness imply  
that a dedicated search program using the image trigger may be required. 
Such search would be useful even in the case of non-detection as this would put 
important constraints on models of star formation in the early Universe, 
models of the GRB progenitors, and the origin of SMBHs.

\section*{Acknowledgments}
This research was funded by STFC under the rolling grant
``Theoretical Astrophysics in Leeds'' (SSK and MVB). 
The work on this project was started during the NORDITA program on Physics of 
Relativistic Flows (Stockholm 2009) and we are grateful to the hosts and organises of 
this program for their generosity and hospitality.
We are also thankful to Felix Aharonyan, Boris Stern, and Alexei Pozanenko for 
useful discussions.

\end{document}